\documentclass[a4paper,11pt]{article}
\pdfoutput=1
\usepackage{jheppub}

\usepackage{slashed}

\newcommand{\BE}{\begin{equation} \begin{array}{c}}
\newcommand{\EE}{\end{array}\end{equation}}

\newcommand{\BT}{\begin{theorem}}
\newcommand{\ET}{\end{theorem}}

\newcommand{\AS}{\slashed{A}}

\newcommand{\LX}{\Lambda}

\newcommand{\lX}{\lambda}

\newcommand{\NN}{\mathbb{N}}
\newcommand{\QQ}{\mathbb{Q}}
\newcommand{\CC}{\mathbb{C}}
\newcommand{\RR}{\mathbb{R}}

\newcommand{\demi}{{\frac{1}{2}}}

\date{\today }

\title {
  A new SU(2/1) supergroup with determinant 1
  explains many mysteries of the weak interactions.
}

\author[a]{Jean Thierry-Mieg}
\affiliation[a]{
  NCBI, National Library of Medicine, National Institute of Health, \\
  8600 Rockville Pike, Bethesda MD20894, U.S.A.
}
\emailAdd{mieg@ncbi.nlm.nih.gov}
\abstract {
  Taken as a classification paradigm completing the standard model, 
  a new compact form of the SU(2/1) supergroup 
  explains many mysterious properties of the weak interactions:
  the maximal breaking of parity,
  the fractional charges of the quarks,
  the cancelation of the quantum field theory anomalies,
  and ties together the existence of the right neutrinos and of the heavier Fermions.
  This compact supergroup is constructed by exponentiating
  the matrices representing the leptons and the quarks
  which form a semi-direct sum of Kac modules
  of the real
  superalgebra $su(2/1,\RR)$ such that the overall
  trace of the $U(1)$ weak-hypercharge $Y$ vanishes.
  Remarkably, all the elements of this supergroup have
  Berezinian 1 and determinant 1.
  In practice, $Tr(Y)=0$ simply means that the
  electric charge of the hydrogen atom is zero.
}
\notoc   % suppress table of content
\begin{document}
\maketitle
\flushbottom
%\tableofcontents

\section{Introduction}
\label{sec:Intro}

The standard model of the strong, weak and electromagnetic interactions is very successful. 
The fundamental forces are fully specified by postulating the $SU(3)SU(2)U(1)$ gauge symmetry
of the Yang-Mills Lagrangian. Yet, the properties of the elementary 
Fermions remain mysterious.
Why do the weak interactions break parity? Why right neutrinos? Why several generations?
Why fractional charges?
Why do quarks and leptons succeed to cancel the quantum field theory (QFT) anomalies?
Unfortunately, neither supersymmetry nor string theory directly addresses any of these fundamental questions.
As a new classification principle completing the standard model, we propose to
consider representations of a new compact form of the $SU(2/1)$ supergroup .

Two pioneers, Ne'eman \cite{N1} and Fairlie \cite{F1}, explored in 1979
the simple Lie-Kac superalgebra $su(2/1)$ \cite{Kac1,Kac2}
because its even subalgebra $su(2)\oplus u(1)$ coincides with 
the electroweak gauge algebra. It was found that the Higgs fields
have the quantum numbers of the odd generators \cite{N1,F1}, 
and that the helicities and weak hypercharges 
of the leptons \cite{N1,F1} and the quarks \cite{DJ,NTM1},
graded by chirality, fit its smallest irreducible representations.
Furthermore the existence of the 3 generations of leptons and quarks becomes 
natural \cite{COQ0,HS98}, because superalgebras
admit indecomposable representations \cite{Marcu80}. 

But all these desirable properties
do no explain the cancellation of the Adler-Bell-Jackiw anomalies \cite{Adler,BJ},
without which gauge invariance is lost and the theory is not renormalizable.
The quarks must, in a  subtle way called the Bouchiat, Iliopoulos and Meyer (BIM) 
mechanism \cite{BIM}, compensate in the quantum loops the deleterious effect of the leptons.
This balance cannot be derived from the Lie algebra mathematics because their 
finite dimensional representations are fully reducible.
Our main objective is to show that a composite set of quartet representations of 
$sl(2/1,\CC)$ with rational Dynkin numbers admits a real form which
exponentiates to a new form of the $SU(2/1)$ supergroup with determinant 1
if and only if the trace of the $U(1)$ weak hypercharge $Y$ vanishes, $Tr(Y)=0$,
and that the corresponding quantum field theory (QFT)
is anomaly free. 

This article is written from the point of view of particle physics but contains several
new results in mathematics. Fixing some confusion in the physics literature, 
we carefully distinguish
the complex superalgebra $sl(2/1,\CC)$, its real form $su(2/1,\RR)$, and the supergroup $SU(2/1)$.
In section \ref{sec:Kac}, all finite dimensional Kac modules \cite {Kac2} of $sl(2/1,\CC)$ are constructed.
In section \ref{sec:Casimir}, the cubic super-Casimir tensor of an $sl(2/1,\CC)$ Kac module is 
shown to be proportional to $Tr(Y)$.
In section \ref{sec:RealForm}, a new generalization of the Berezin super-Hermitian conjugation 
and new representations of $su(2/1,\RR)$ are defined.
In section \ref{sec:Mat} new  nested indecomposable representations of $su(2/1,\RR)$ are constructed.
In section \ref{sec:Group}, a new form of the $SU(2/1)$ supergroup is constructed 
by exponentiation and shown to be compact and to have unit Berezinian and unit determinant
if $Tr(Y)=0$.
In section \ref{sec:QFT}, 
it is shown that the even sector of the super-Casimir tensor provides an algebraic meaning to
the quantum field theory anomalies \cite{Adler,BJ}.
Finally, in section \ref{sec:SM}, it is shown that the BIM composite module,
3 quarks per lepton, satisfies the selection rule $Tr(Y)=0$.
We therefore conclude that 
the vanishing of the trace of the weak hypercharge $Y$ 
implies at the same time a supergroup with determinant 1 
and the consistency of the quantum field theory.

\section {The $sl(2/1,\CC)$ Kac modules}
\label{sec:Kac}

The simple Lie algebra $sl(2,\CC)$ admits 3 Chevalley generators $(f,h,e)$
with commutators
\BE
\label{eq:Kac.1}
[h,e]=2e\;,\;\;[h,f]=-2f\;,\;\;[e,f]=h\;.
\EE
A finite representation of dimension $(a+1)$ is provided by the iterated action of $f$ on a highest weight state $|a>$,
where $a$ is the non-negative integer Dynkin number of the representation, modulo the vector space
generated by $\{f^{a+i}|a>,i>0\}$.
\BE
\label{eq:Kac.2}
h|a-2i>=(a-2i)|a-2i>\;,\;\;i=0,1,2 \dots ,a\\
e|a>=f|-a>=0\;,\\
f|a-2i> = |a-2i-2>\;,\;\;
e|a-2i-2>=(i+1)(a-i)|a-2i>\;,\;\;i=0,\dots,a-1
\EE
To construct the Lie superalgebra $sl(2/1,\CC)$, add a Cartan generator $k$ and a pair
of anticommuting odd generators $(u,v)$ satisfying
\BE
\label{eq:Kac.3}
[h,u]=-u,\;\{u,v\}=k,\;[k,e]=e,\;\;[k,h]=[k,u]=[e,[e,u]]=u^2=v^2=0.
\EE
A Kac module \cite{Kac2} is provided by the iterated action of $(f,v)$ on a highest
weight state $\LX_0 = |a,b>$ where $a$ is a non negative integer and $b$ is complex:
\BE
\label{eq:Kac.4}
\LX_0=|a,b>\;,\;\;e\LX_0=u\LX_0=0\;,\;\;h\LX_0=a\LX_0\;,\;\;k\LX_0=b\LX_0\;,
\EE
modulo the $a$-negative even submodules.
Since $v^2=0$, the complex number $b$, called the odd Dynkin weight,
is not quantized: if $a \in \NN$, $\forall b \in \CC$, the representation has finite dimension $4(a+1)$. Using (\ref{eq:Kac.1}-\ref{eq:Kac.4}), if $a>0$, the four weights 
\BE
\label{eq:Kac.5}
\LX_0,\;\;\;\LX_1=v\LX_0,\;\;\;\LX_2=(fv-(a+1)vf/a)\LX_0,\;\;\LX_3=vfv\LX_0,
\EE
are highest weight of the even subalgebra: $e \LX_i = 0,\;\forall i=0,1,2,3$.
If $a=0$, $\LX_2$ is removed. The generator $Y=2k - h$ is called the hypercharge.
It commutes with $(f,h,e)$ and generates the center $\CC$ of the even subalgebra $sl(2)\oplus \CC$.
On the 4 even submodules, $Y$ has eigenvalues $(y=2b-a, y-1, y-1, y-2)$.
The adjoint representation
$a=b=1$ and all other Kac modules satisfying $y=2b-a=1$ are real in the sense that they are invariant
if we flip the signs of $(h,k)$, exchange $(e,f)$ and map $(u,v)$ to $(iv,iu)$.

There are 2 special cases, called atypical \cite{Kac2}. Since $u \LX_1 = b \LX_0$, if $b=0$ then $\LX_1$
is an $sl(2/1)$ highest weight and the Kac module is indecomposable rather than irreducible.
This case is called atypical 1. Similarly, since $(eu-ue)\LX_3 = (b-a-1) \LX_1$,
if $b=a+1,a>0$ then $\LX_2$ (or $\LX_3$ if $b=1,a=0$) is an $sl(2/1)$ highest weight
and this case is called atypical 2.
Yet, these states are not quotiented out of the Kac module
and the total dimension remains $4(a+1)$.

This formal study defines the matrices $\forall\;a,b$, and will be used in section \ref{sec:Casimir}.
To better understand the situation and for the application to particle physics, it is
helpful to visualize the 8 generators $\lX$ in the case $a=0$, $y=2b$.
This fundamental quartet was first constructed by Nahm, Scheunert and Rittenberg in 1977 \cite{SNR77}.
The 4 even matrices are
\BE
\label{eq:Kac.6}
f = \begin{pmatrix} 
 0 & 0 & 0 & 0 \cr 0 & 0 & 0 & 0 \cr 0 & 1 & 0 & 0 
 \cr 0 & 0 & 0 & 0
\end{pmatrix}
\;,\;\;\;
h= \begin{pmatrix} 
      0 & 0 & 0 & 0
  \cr 0 & 1 & 0 & 0
  \cr 0 & 0 & -1 & 0 
  \cr 0 & 0 & 0 & 0 
\end{pmatrix}
,\;\;\;
e = \begin{pmatrix} 
 0 & 0 & 0 & 0 \cr 0 & 0 & 1 & 0 \cr 0 & 0 & 0 & 0 
 \cr 0 & 0 & 0 & 0
\end{pmatrix}
,\;\;\;
Y = \begin{pmatrix} 
 y & 0 & 0 & 0 \cr 0 & y-1 & 0 & 0 \cr 0 & 0 & y-1 & 0 
 \cr 0 & 0 & 0 & y-2 
\end{pmatrix}
,
\EE
$(f,e,h)$ generate $sl(2,\CC)$ and commute with $Y$.  $u,v,w=[e,u],x=-[f,v]$ are odd:
\BE
\label{eq:Kac.7}
u = \begin{pmatrix} 
      0 & b & 0 & 0 
  \cr 0 & 0 & 0 & 0
  \cr 0 & 0 & 0 & b-1
  \cr 0 & 0 & 0 & 0 
\end{pmatrix}
,\;\;\;
v = \begin{pmatrix} 
      0 & 0 & 0 & 0 
  \cr 1 & 0 & 0 & 0
  \cr 0 & 0 & 0 & 0
  \cr 0 & 0 & 1 & 0
\end{pmatrix}
,\;\;\;
w = \begin{pmatrix} 
      0 & 0 & -b & 0
  \cr 0 & 0 & 0 & b-1
  \cr 0 & 0 & 0 & 0
  \cr 0 & 0 & 0 & 0 
\end{pmatrix}
,\;\;\;
x = \begin{pmatrix}   
      0 & 0 & 0 & 0
  \cr 0 & 0 & 0 & 0
  \cr -1 & 0 & 0 & 0
  \cr 0 & 1 & 0 & 0
\end{pmatrix}
.
\EE
They form by commutation with $(f,h,e,Y)$ a complex doublet of $sl(2)\oplus \CC$ and close by
anticommutation on the even matrices. The indecomposable structure if $b=0$ or $b=1$ is very visible.
In addition, the superidentity $\chi = diag(-1,1,1,-1)$ defines the supertrace and the grading. $\chi$ commutes with
the even generators $\lX_a$ and anticommutes with the odd generators $\lX_i$
\BE
\label{eq:Kac.8}
\chi = diag(-1;1,1;-1)\;,\;\;STr(...) = Tr(\chi\;...)\;,
\\
\;\;[\chi,\lX_a] = \{\chi,\lX_i\}=0\;,\;\;\lX_a=f,h,e,y,\;\;\lX_i=u,v,w,x.
\EE
All these matrices have vanishing supertrace, they also all have vanishing trace except the hypercharge
for which $Tr(Y)=4(y-1)$.
This crucial detail will play a fundamental role in our analysis.

\section {The $sl(2/1,\CC)$ cubic super-Casimir tensor is proportional to $Tr(Y)$}
\label{sec:Casimir}

The cubic super-Casimir tensor $C$ is defined as the supertrace of the correctly symmetrized
products of 3 matrices. There are 2 non trivial sectors, even-even-even and even-odd-odd,
or rather 5 sectors if the hypercharge $Y$ is distinguished from the $sl(2,\CC)$ generators
$(\lX_a=f,h,e)$ and from the odd generators $(\lX_i=u,v,x,y)$.
\BE
\label{eq:Casimir.1}
C_{abc} = \frac{1}{6}\;STr (\lX_a\;\{\lX_b,\lX_c\})\;,\;\; 
C_{yab} = \frac{1}{2}\;STr (Y\;\{\lX_a,\lX_b\})\;,\;\; 
C_{yyy} = STr (Y^3)
\\
C_{aij} = \frac{1}{2}\;STr (\lX_a\;[\lX_i,\lX_j])\;,\;\; 
C_{yij} = \frac{1}{2}\;STr (Y\;[\lX_i,\lX_j])\;.
\EE
To evaluate these supertraces over a Kac module 
with highest weight $(a \in \NN,y \in \CC)$,
first compute the traces over its 4 submodules $\{\LX_i,i=1.2.3.4\}$ (\ref{eq:Kac.5})
with highest weights $(a',y')=\{(a,y),(a+1,y-1),(a-1,y-1),(a,y-2)\}$
and super-dimensions $\{-(a+1),a,a+2,-(a+1)\}$. Then add up.
By inspection $Tr(Y)=4(a+1)(y-1)$. 

In each even sector $C_{abc}(a',y')=0$ because $su(2)$ does not admit a cubic-Casimir tensor.
By inspection $C_{yab}(a',y')=a'(a'+1)(a'+2)y'/3$, and $C_{yyy}(a',y')=(a'+1)y'^3$. 
Super-adding the 4 sectors with signs $(-1,1,1,-1)$ then gives
\BE
\label{eq:Casimir.2}
C_{yab} = 2(a+1)(y-1)\delta_{ab}=\frac{1}{2}\delta_{ab}\;\;Tr(Y)\;,\;\; 
C_{yyy} = -6(a+1)(y-1)=-\frac{3}{2}\;Tr(Y)\;,\;\; 
\EE
The odd sector can be evaluated using the definitions (\ref{eq:Kac.1}-\ref{eq:Kac.4}),
with patience or a computer, on enough examples \cite{TMJG22} to saturate a polynomial of degree at most 4 in $(a,y)$, giving
\BE
\label{eq:Casimir.3}
C_{yuv} = \frac{1}{2} STr (Y\;[u,v]) = -\frac{3}{8}\;Tr(Y)\;,\;\; 
C_{huv} = \frac{1}{2} STr (h\;[u,v]) = -\frac{1}{8}\;Tr(Y)\;.\;\; 
\EE
In conclusion, all components of the $sl(2/1,\CC)$ cubic super-Casimir tensor super-traced
over a Kac module are linear in $Tr(Y)=4(a+1)(y-1)$
with no other dependence in the Dynkin numbers $(a,b)$.
This result also applies to the trace metric and to 
several other trace tensors
appearing in the calculation of the Feynman diagrams of
the scalar-vector-tensor super-chiral quantum field theory \cite{TM20b,TMJ21a}
\BE
\label{eq:Casimir.4}
Tr(uv) = \frac{1}{4}\;Tr(Y)\;,\;\;\;
Tr(uvwx) = \frac{1}{4}\;Tr(Y)\;,\\
Tr(Y \;[u,v]) = \frac{1}{4}\;Tr(Y)\;,\;\;\;
Tr(h \;[u,v]) = -\frac{1}{4}\;Tr(Y)\;.
\EE

\section{The $su(2/1,\RR)$ real form}
\label{sec:RealForm}

By introducing an appropriate phase co-factor $\eta = exp(i\pi/4)$ and by symmetrization,
one can extract from any Kac module representation of $sl(2/1,\CC)$ with real Dynkin numbers, a set
of zeta-Hermitian matrices
\BE
\label{eq:RF.1}
  \lX = \lX^{\#} = \zeta \lX^{\dagger}\zeta\;,
\EE
where $\zeta$ is a real diagonal matrix, commuting with the even subalgebra 
and satisfying $\zeta^2= Id$. The signs of the  $\pm1$ eigenvalues along the
diagonal are controlled by the signs of the Kac atypicality conditions $(b)$ and $(b-a-1)$.
This construction is valid for all typical or singly-atypical $sl(m/n)$ Kac modules (in preparation).
As an example, we give the $\zeta$ matrix and the odd matrices 
of the $a=0$ Kac module. We keep (\ref{eq:Kac.6}) and modify (\ref{eq:Kac.7}):
\BE
\label{eq:RF.2}
\zeta = \begin{pmatrix} 
      1 & 0 & 0 & 0
  \cr 0 & s & 0 & 0
  \cr 0 & 0 & s & 0
  \cr 0 & 0 & 0 & st 
\end{pmatrix}
\;,\;\;\;
\lX_4 = \begin{pmatrix} 
      0 & \beta & 0 & 0
  \cr \beta & 0 & 0 & 0
  \cr 0 & 0 & 0 & \gamma
  \cr 0 & 0  & \gamma & 0
\end{pmatrix}
\;,\;\;\;
\lX_6 = \begin{pmatrix} 
      0 & 0 -\beta & 0 
  \cr 0 & 0 & 0 & \gamma
  \cr -\beta & 0 & 0 & 0
  \cr 0 & \gamma & 0 & 0
\end{pmatrix}
\;,
\\
\\
\lX_5 = \begin{pmatrix} 
      0 & -i\beta & 0 & 0
  \cr i\beta & 0 & 0 & 0
  \cr 0 & 0 & 0 & -i\gamma
  \cr 0 & 0  & i\gamma & 0
\end{pmatrix}
\;,\;\;\;
\lX_7 = \begin{pmatrix} 
      0 & 0 & i\beta & 0 
  \cr 0 & 0 & 0 & -i\gamma
  \cr -i\beta & 0 & 0 & 0
  \cr 0 & i\gamma & 0 & 0
\end{pmatrix}
\;,
\\
\\
s=sign(b),\;t=sign(b-1),\;\;\;\;\beta=\sqrt{b},\;\gamma=\sqrt{b-1}\,.
\EE
One can then construct `anti`-$\zeta$-Hermitian matrices
\BE
\label{eq:RF.3}
\mu_a = i \lX_a\;,\;\; \mu_a=-\mu_a^{\dagger}= -\mu_a^{\#}\;,
\\
\mu_j = \eta \lX_j\;,\;\;\;\mu_i = i \mu_i^{\#} \;,
\EE
where $(\lX_a,\lX_i)$ are respectively symmetrized forms of $(f,h,e,k)$ and $(\lX_4,\lX_5,\lX_6,\lX_7)$
The even $\mu_a$ matrices represent $su(2,\RR)$. By inspection,
the odd $\mu$ matrices close under super-commutation using real structure constants.
Therefore, the $\mu$ matrices represent the real superalgebra $su(2/1,\RR)$.
The co-factor $\eta$ was introduced in the case $a=0$ by Furutsu \cite{Furutsu88,
Furutsu89}, but he wrongly concluded that his construction of a real form
of $sl(2/1,\CC)$ was only valid for $b>1$ or $b<0$. 
Indeed, if $b>1$, the $\zeta$  matrix is the identity: $\zeta=Id$
and the $\lX$ matrices are all Hermitian: $\lX^{\#}=\lX^{\dagger}$. 
If $b<0$, the $\zeta$  matrix (\ref{eq:RF.2}) coincides, up to a sign choice, with
the super-identity  $\zeta=-\chi$ (\ref{eq:Kac.8}), and the $\lX$ matrices are super-Hermitian
in the sense of Berezin: $\lX^{\#}=\chi \lX^{\dagger} \chi$. 
But no one realized that when $0<b<a+1$ one could
interpolate between $Id$ and $\chi$ of Berezin and define $\zeta$-Hermitian matrices:
$\lX^{\#}=\zeta \lX^{\dagger} \zeta$. 

\section {Nested indecomposable matryoshka $su(2/1,\RR)$ representations}
\label{sec:Mat}
Let $\lX$ be as above a set of square matrices of size $4(a+1)$ representing an $sl(2/1,\CC)$ 
or $su(2/1,\RR)$ Kac module
with highest weight $|a,b>,y=2b-a$.
Let $N$ be a non negative integer. Construct a banded lower triangular block matrix $\LX$ of size $4N(a+1)$ such that
the $n$-$th$ column coincides with the first $N - n + 1$ terms of the Taylor development of $\lX(y)$ relative to $y$. For example if $N=3$
\BE
\label{eq:Mat.1}
\LX = \begin{pmatrix} 
      \lX & 0 & 0 
  \cr \lX' & \lX & 0 
  \cr \demi{{\lX}^{''}}  & \lX' & \lX 
\end{pmatrix}
\;.
\EE
By construction, the matrices $\LX$ have the same commutation rules as the matrices $\lX$ because the
commutators of the Taylor expansion coincide with the Leibniz development of the derivatives of the commutators $[a,b]^{''}= [a^{''},b] + 2 [a',b'] + [a,b^{''}]$.
In a Lie algebra, the Dynkin numbers are integral, such derivatives cannot be computed, and all finite dimensional representations are fully reducible. But in $sl(2/1)$,
$y$ is not quantized and the matrices $\LX$ generate recursively nested indecomposable representations where each
generations is coupled via mouse trap coefficients $(\lX',\lX'',\dots)$ to the upper generations.
Notice that the even matrices $(f,h,e)$ representing $sl(2)$ are block diagonal, because their elements (\ref{eq:Kac.2})
are independent of $y$. The hypercharge only contributes to the main and second diagonal since $Y$ is linear in $y$ (\ref{eq:Kac.4}).
However, in the $su(2/1,\RR)$ $\zeta$-Hermitian case (\ref{eq:RF.2}), the $n$-$th$ $y$-derivatives of the odd matrices $(\mu_i,\;i=4,5,6,7)$ do not vanish and fill 
the whole lower triangle. Such mixing is specific of superalgebras \cite{Todorov23}. 

We call these representations matryoshkas because they are recursively nested like
Russian dolls.
The $N=3$ representation of $sl(2/1,\CC)$ was first constructed by Marcu \cite{Marcu80}.
Germoni proved that these are the only indecomposable representations of $sl(2/1)$ corresponding
to semi-direct sums of Kac modules \cite{Germoni98}. With Jarvis, Germoni and Gorelik, we proved recently \cite{TMJG23} that such
representations exist for any $N$ and any Kac module of a type 1 superalgebra, the series $sl(m/n,\CC)$ and $osp(2/2n,\CC)$, but did not study $su(2/1,\RR)$.

\section{The 'super-special' SU(2/1) supergroup with determinant 1}
\label{sec:Group}

Consider a quartet representation $\mu$ of $su(2/1,\RR)$ (\ref{eq:RF.3}). 
The exponential $exp(x^a\mu_a)$ of the anti-Hermitian even matrices $\mu_a,\;a=1,2,3$ 
is periodic in $x^a$.
To ensure the existence of a single neutral element, $x$ must be
quotiented modulo the period. If all $su(2)$  Dynkin number are even, the period is $\pi$
giving the compact Lie group $SO(3)$, otherwise the period is $2\pi$ giving 
its double cover $SU(2)$.
If and only if $y=p/q$ is rational, $exp(ixY)$ with $Y$ eigenvalues $(y,y-1,y-2)$ 
(\ref{eq:Kac.6}) is periodic with period $2 q \pi$.
Modulo the period, $iY$ exponentiates to the compact $U(1)$ group.
Henceif $y \in \QQ$ the exponential map of the even generators gives the $SU(2)U(1)$ compact group.

The exponential of the odd generators $\mu_i=\eta\lX_i,\;i=4,5,6,7$ (\ref{eq:RF.2}) 
using 4 distinct anticommuting 
Grassmann parameters $\theta^i$ \cite{Nieto93} is discrete and finite.
\BE
\label{eq:Group.1}
X = \theta^i\mu_i\;,\;\;e^X = 1 + X + \frac{1}{2}X^2 + \frac{1}{6}X^3 + \frac{1}{24}X^4\;.
\EE
Hence the whole $SU(2/1)$ supergroup is compact. Because of the presence of Grassmann parameters,
the Berezinian is multiplicative $Ber(MN)=Ber(M)Ber(N)$, but the determinant is not
$det(MN) \neq det(M)det(N)$. However, in the quartet case (\ref{eq:RF.2},\ref{eq:RF.3}) a direct calculation
respecting the order of the columns shows that:
\BE
\label{eq:Group.2}
Ber(e^X) = e^{STr(X)} = e^0 = 1
\;,\\
det(e^X) = 1 - \frac{1}{2}(\theta^4\theta^5 + \theta^6\theta^7)\;Tr(Y) - \frac{5}{3}\theta^4\theta^5\theta^6\theta^7\;Tr(Y)
\;.
\EE
Since the determinant is linear in $Tr(Y)$, the determinant of the exponential map of
a direct sum of quartets, i.e. of a set of block diagonal matrices, acts additively mimicking
the even exponential map $det(e^{iY})= e^{i Tr(Y)}$.
Therefore if $Y$ traced over a composite Kac module vanishes
$Tr(Y)=0$, the exponential map has determinant 1 
\BE
\label{eq:Group.3}
Tr(Y) = 0 \Rightarrow \forall g=exp(x^a\mu_a+\theta^i\mu_i),\;det(g)=Ber(g)=1.
\EE
We propose to call the group $\{g\}$ the  'super-special' $SU(2/1)$ supergroup.

The key of this construction is the linearity of $e^X$ in $Tr(Y)$. 
Our proof holds for the fundamental Kac quartet $(a=0,\;b \in \QQ)$,
and for their semi-direct indecomposable sums (\ref{eq:Mat.1}) as
lower triangular elements do not contribute to determinants.
In the shited adjoint case $(a=1)$ \cite{TMJG22}, $det(g)$ in optimal column order is also linear in $Tr(Y)$.
An open problem is to extend the proof to all Kac modules $(a \in \NN,\; b \in \QQ)$.

The vanishing of the term in $\theta^2$ in (\ref{eq:Group.2}) matches the vanishing of the anomalous
trace term of the tensor propagator of the $su(2/1)$ super-chiral model (eq.(3.7) in \cite{TMJ21a}).
The vanishing of the term in $\theta^4$ in (\ref{eq:Group.2}) is related to the vanishing of the anomalous
trace terms of the Feynman diagrams with 4 external legs (eq.(2.12) in \cite{TM20b}).

\section{Quantum field theory anomaly cancellation}
\label{sec:QFT}
Consider a set of massless Fermions belonging to a semi-direct sum of
Kac modules of the superalgebra $su(2/1,\RR)$ graded
by chirality. That is assume that the superidentity grading operator 
$\chi$ of $su(2/1)$ (\ref{eq:Kac.8}) has eigenvalue
$+1$ on the left Fermion states and $-1$ on the right Fermion states. Add in $su(2)\oplus u(1)$ Yang-Mills vectors coupled
to the Fermions using the even matrices $\mu_a$ (\ref{eq:RF.3}) of $su(2/1,\RR)$: $\AS=\AS^a \mu_a$. Because such Fermions are graded by chirality, the
supertrace in the sense of the superalgebra $STr(...)=Tr(\chi\;...)$ is equivalent to
the $\gamma^5$ trace $\Gamma Tr(...)=Tr(\gamma^5\;...)$ which occurs when computing 
the Adler-Bell-Jackiw \cite{Adler,BJ}  triangle
anomaly  $d_{abc} = \Gamma Tr (\mu_a \{\mu_b ,\mu_c\})$. This presentation
provides an algebraic meaning to the anomaly tensor $d_{abc}$
which now coincides with the even sector of the cubic super-Casimir tensor of the real simple superalgebra $su(2/1,\RR)$.
As shown above, if $Tr(Y)=0$, the super-Casimir tensor vanishes,
hence the $su(2)^2Y$ and $Y^3$ tensors both vanish (\ref{eq:Casimir.2})
and the theory is anomaly free.

If the $SU(3)$ color group gauging the strong interactions is added to this construction,
there is a further potential anomaly $su(3)^2Y$. But since $su(3)$ commutes
with $su(2/1)$, the $su(3)^2Y$  anomaly factorizes and vanishes because $STr(Y)=0$ on any $su(2/1)$ representation.

\section {Applying  the $SU(2/1)$ supergroup to the standard model}
\label{sec:SM}
The electron and neutrino \cite{N1,F1} graded by chirality $(\nu_R;\nu_L,e_L;e_R)$
have the quantum numbers of the fundamental Kac module $(a=0,y=0)$ of
$su(2/1,\RR)$ (\ref{eq:Kac.6})
with electric charge $Q=(Y-h)/2$.
The top singlet $(0,0)$ represents the right neutrino $\nu_R$.
The quarks \cite{DJ,NTM1} graded by chirality $(u_R;u_L,d_L;d_R)$ have the quantum numbers of
the irreducible typical $(a=0,y=4/3)$ $su(2/1,\RR)$ representation (\ref{eq:RF.3}).
In both cases, $Tr(Y)=4(y-1)$, respectively $-4$ for the leptons and $+4/3$ for the quarks.
Therefore, as there exist 3 quark colors in each lepton family, $Tr(Y)=0$,
the theory is anomaly free (section \ref{sec:QFT}), and the hydrogen atom $(e^{-1},u^{2/3},u^{2/3},d^{-1/3})$ has
electric charge $Q=0$.

We have thus discovered a new implication of the celebrated BIM mechanism \cite{BIM},
which allowed to predict the existence of the charm and the top quarks
as soon as the strange and bottom quarks were recognized:
the BIM mechanism also allows exponentiation to a new form of the 
$SU(2/1)$ supergroup with determinant one (section \ref{sec:Group}).

Furthermore, the 3 generations of leptons, and separately the 3 generations of quarks,
can each be grouped into a single 12 dimensional 
nested indecomposable representation \cite{TMJG23}.
Since the $\lX',\lX'',\dots$ blocks of the $\LX$ matrices  (\ref{eq:Mat.1})
are below the main diagonal, the 3-generations model remains anomaly free.
Cherry on the cake, the existence of three generations of charged leptons implies the existence of
three generations of right neutrinos $\nu_R$, because without them the remaining atypical irreducible
fundamental lepton triplet $(\nu_L,e_L/e_R)$
considered in \cite{N1,F1}
is quantized ($y=0$), the $\lX'$ derivatives cannot be computed, and 
the 3 generations indecomposable representation $\LX$ cannot be constructed.
This impossibility can also be proved by cohomology \cite{JTM22}.

\section{Conclusion}
\label{sec:ccn}
Considered as a classification paradigm, the $SU(2/1)$ supergroup with determinant $1$
constructed in section \ref{sec:Group}
answers all the questions stated in the introduction.

The basic observation is that the $SU(2)U(1)$ even subgroup coincides with the electroweak 
gauge group of  the standard model.
The leptons and the quarks fit the fundamental representation (\ref{eq:Kac.6},\ref{eq:RF.2})
of the real $su(2/1,\RR)$ superalgebra \cite{N1,F1,DJ,NTM1},
but since all these particles are chiral Fermions, it is natural to grade
$su(2/1)$ by chirality, relating left states to right states, rather than
relating Bosons to Fermions as in traditional supersymmetric models.
Then, thanks to mathematics, the desired results follow.

Why is parity broken? Maximal parity breaking, the deepest mystery of the weak interactions,
is implied by the grading. The fundamental $su(2/1)$ quartet $(1;2;1)$  
is asymmetric (\ref{eq:Kac.6}). The weak $su(2)$ algebra only interacts with the central doublet
and ignores the singlets. Therefore, grading by chirality $(1_R;2_L;1_R)$ fully breaks parity.

Why fractional charges? As observed
experimentally, the $u(1)$ weak hypercharge $Y$ (\ref{eq:Kac.6}) steps by full units $(y;y-1,y-1;y-2)$,
so the righ
t neutrino $\nu^e_R$ is neutral $(y=0)$  because the electron $e^-$ and the weak vector Boson
$W^-_{\mu}$ have the same
electric charge $(-1)$ if and only if $(y=0)$.
Yet $Y$ is not quantized and the fractional charge $y=4/3$ of the $u_R$ quark
is allowed (\ref{eq:Kac.6}). Our new generalization of the Berezin super-conjugation (\ref{eq:RF.1})
then provides a real form of the quark representation (\ref{eq:RF.3}), a required property since the photon is its own anti-particle,
hence the $u(1)$ subalgebra must be real.

Why several generations, why right neutrinos? The existence of several generations of leptons and quarks,
a hard problem \cite{Todorov23},
becomes natural because $su(2/1)$ admits (\ref{eq:RF.3},\ref{eq:Mat.1})
 nested indecomposable matryoshka
representations \cite{Marcu80,COQ0,HS98,TMJG23}. In turn, section \ref{sec:SM},
the existence of the three generations of heavier Fermions implies the
existence of the three right neutrinos, because the irreducible triplet $(\nu_L,e_L;e_R)$
does not admit a multi-generations indecomposable extension \cite{JTM22}.

Why do anomalies cancel out? The vanishing of the Adler-Bell-Jackiw anomalies,
and this is probably the most striking result of this study, 
is implied because this form of the $SU(2/1)$ supergroup has determinant 1.
A simple calculation (\ref{eq:Casimir.2}) shows that the symmetric tensor $d_{\{abc\}}$
traced over a semi-direct sum of Kac modules vanishes,
$Tr(\LX_a\;\{\LX_b,\LX_c\})=0$, if and only if $Tr(Y)=0$.
We then observed  (\ref{eq:Group.3}) that the superalgebra matrices (\ref{eq:Kac.6},\ref{eq:RF.2})
exponentiate to a matrix group with determinant 1 if and only if $Tr(Y)=0$.
But since the superalgebra is
graded by chirality, section \ref{sec:QFT}, its supertrace coincides with 
the $\gamma_5$ trace appearing in quantum field theory
when evaluating the anomalies. Therefore, exponentiation to the 'super-special' $SU(2/1)$ supergroup
is equivalent to the vanishing of the anomalies and both follow from the simple selection rule $Tr(Y)=0$. 

Guided by particle physics, we computed en passant new tensor identities (\ref{eq:Casimir.2},\ref{eq:Casimir.4}),
defined new  $\zeta$-Hermitian extensions (\ref{eq:RF.1}) of the Berezin super-transposition,
found contrary to the literature new representations (\ref{eq:RF.1}) of the real superalgebra (\ref{eq:RF.2}) with hypercharge $0<y<1$,
constructed in a simple way  (\ref{eq:Mat.1}) new nested indecomposable representations of $sl(2,1,\RR)$, 
and discovered the surprising existence,
not reported in the recent review of Fioresi and Gavarini \cite{Fioresi23} or in Saleur and Schomerus \cite{Saleur07}, of a 
super-special supergroup  (\ref{eq:Group.3}) whose elements
have determinant and super-determinant (Berezinian) equal to 1. This double structure might be
related to the existence of two parallel transports on a supergroup, via the adjoint and via the
alternative adjoint action of Arnaudon, Bauer and Frappat \cite{Arnaudon97}, and to their ghost Casimir 
generalized by Gorelik \cite{Gorelik2000}. We conjecture that
both transports are consistent if and only if the corresponding quantum field theory is anomaly free.

Thanks to these new results, one can, for the first time, regard the electroweak interactions
as structured by the $SU(2/1)$ supergroup.
\acknowledgments

We gratefully acknowledge Joris van der Jeugt, Maria Gorelik, Victor Kac, Peter Jarvis and Danielle Thierry-Mieg for crucial comments.
This research was supported by the Intramural Research Program of the National Library of Medicine, National Institute of Health.

%% The Appendices part is started with the command \appendix;
%% appendix sections are then done as normal sections
\appendix

%\nocite{*}
\bibliographystyle{unsrt}
\bibliography{Yzero}

\begin{thebibliography}{10}

\bibitem{N1}
Yuval Ne'eman.
\newblock Irreducible gauge theory of a consolidated {S}alam-{W}einberg model.
\newblock {\em Physics Letters B}, 81(2):190--194, 1979.

\bibitem{F1}
DB~Fairlie.
\newblock Higgs fields and the determination of the {W}einberg angle.
\newblock {\em Physics Letters B}, 82(1):97--100, 1979.

\bibitem{Kac1}
Victor~G Kac.
\newblock Lie superalgebras.
\newblock {\em Advances in mathematics}, 26(1):8--96, 1977.

\bibitem{Kac2}
VG~Kac.
\newblock Characters of typical representations of classical {L}ie
  superalgebras.
\newblock {\em Communications in Algebra}, 5(8):889--897, 1977.

\bibitem{DJ}
P.H. Dondi and P.D. Jarvis.
\newblock A supersymmetric {W}einberg-{S}alam model.
\newblock {\em Physics Letters B}, 84(1):75--78, 1979.

\bibitem{NTM1}
Yuval Ne'eman and Jean Thierry-Mieg.
\newblock Geometrical gauge theory of ghost and goldstone fields and of ghost
  symmetries.
\newblock {\em Proceedings of the National Academy of Sciences},
  77(2):720--723, 1980.

\bibitem{COQ0}
Robert Coquereaux.
\newblock Elementary fermions and su(2/1) representations.
\newblock {\em Physics Letters B}, 261(4):449--458, 1991.

\bibitem{HS98}
Rainer H{\"a}ussling and Florian Scheck.
\newblock Triangular mass matrices of quarks and
  {C}abibbo-{K}obayashi-{M}askawa mixing.
\newblock {\em Physical Review D}, 57(11):6656, 1998.

\bibitem{Marcu80}
Mihael Marcu.
\newblock The representations of spl(2,1)—an example of representations of
  basic superalgebras.
\newblock {\em Journal of Mathematical Physics}, 21(6):1277--1283, 1980.

\bibitem{Adler}
Stephen~L Adler.
\newblock Axial-vector vertex in spinor electrodynamics.
\newblock {\em Physical Review}, 177(5):2426, 1969.

\bibitem{BJ}
John~Stewart Bell and Roman~W Jackiw.
\newblock A {PCAC} puzzle.
\newblock {\em Nuovo cimento}, 60(CERN-TH-920):47--61, 1969.

\bibitem{BIM}
C.~Bouchiat, J.~Iliopoulos, and Ph. Meyer.
\newblock An anomaly-free version of {W}einberg's model.
\newblock {\em Physics Letters B}, 38(7):519--523, 1972.

\bibitem{SNR77}
Manfred Scheunert, Werner Nahm, and Vladimir Rittenberg.
\newblock Irreducible representations of the osp (2,1) and spl (2,1) graded
  {L}ie algebras.
\newblock {\em Journal of Mathematical Physics}, 18(1):155--162, 1977.

\bibitem{TMJG22}
Jean Thierry-Mieg, Peter~D. Jarvis, and Jerome Germoni.
\newblock Explicit construction of the finite dimensional indecomposable
  representations of the simple lie-kac $su(2/1)$ superalgebra and their low
  level non diagonal super casimir operators.
\newblock {\em arxiv:2207.06545}, 2022.

\bibitem{TM20b}
Jean Thierry-Mieg.
\newblock Scalar anomaly cancellation reveals the hidden superalgebraic
  structure of the quantum chiral {SU}(2/1) model of leptons and quarks.
\newblock {\em Journal of High Energy Physics}, 2020(10), oct 2020.

\bibitem{TMJ21a}
Jean Thierry-Mieg and Peter Jarvis.
\newblock {SU} (2/1) superchiral self-duality: a new quantum, algebraic and
  geometric paradigm to describe the electroweak interactions.
\newblock {\em Journal of High Energy Physics}, 2021(4):1--25, 2021.

\bibitem{Furutsu88}
Hirotoshi Furutsu and Takeshi Hirai.
\newblock Representations of {L}ie superalgebras i extensions of
  representations of the even part.
\newblock {\em Journal of mathematics of Kyoto University}, 28(4):695--749,
  1988.

\bibitem{Furutsu89}
Hirotoshi Furutsu.
\newblock Representations of {L}ie superalgebras, ii unitary representations of
  {L}ie superalgebras of type a(n, 0).
\newblock {\em Journal of Mathematics of Kyoto University}, 29(4):671--687,
  1989.

\bibitem{Todorov23}
Ivan Todorov.
\newblock Octonion internal space algebra for the standard model.
\newblock {\em Universe}, 9(5):222, 2023.

\bibitem{Germoni98}
Jerome Germoni.
\newblock Indecomposable representations of special linear {L}ie superalgebras.
\newblock {\em Journal of Algebra}, 209(2):367--401, 1998.

\bibitem{TMJG23}
Jean Thierry-Mieg, Peter~D. Jarvis, Jerome Germoni, and Maria Gorelik.
\newblock Construction of matryoshka nested indecomposable n-replications of
  {K}ac-modules of quasi-reductive {L}ie superalgebras, including the sl(m/n)
  and osp(2/2n) series.
\newblock {\em SciPost Phys. Proc.}, page 045, 2023.

\bibitem{Nieto93}
V~Hussin and LM~Nieto.
\newblock Supergroups factorizations through matrix realization.
\newblock {\em Journal of mathematical physics}, 34(9):4199--4220, 1993.

\bibitem{JTM22}
Peter~D Jarvis and Jean Thierry-Mieg.
\newblock Indecomposable doubling for representations of the type i lie
  superalgebras sl (m/n) and osp (2/2n).
\newblock {\em Journal of Physics A: Mathematical and Theoretical},
  55(47):475206, 2022.

\bibitem{Fioresi23}
Rita Fioresi and Fabio Gavarini.
\newblock Real forms of complex lie superalgebras and supergroups.
\newblock {\em Communications in Mathematical Physics}, 397(2):937--965, 2023.

\bibitem{Saleur07}
Hubert Saleur and Volker Schomerus.
\newblock On the su(2|1) wznw model and its statistical mechanics applications.
\newblock {\em Nuclear Physics B}, 775(3):312--340, 2007.

\bibitem{Arnaudon97}
Daniel Arnaudon, Michel Bauer, and L~Frappat.
\newblock On {C}asimir's ghost.
\newblock {\em Communications in mathematical physics}, 187(2):429--439, 1997.

\bibitem{Gorelik2000}
Maria Gorelik.
\newblock On the ghost centre of {L}ie superalgebras.
\newblock In {\em Annales de l'institut Fourier}, volume 50/6, pages
  1745--1764, 2000.

\end{thebibliography}
\end{document}